\input harvmac
\input epsf
\input amssym
%
%
\noblackbox
\newcount\figno
\figno=0
\def\fig#1#2#3{
\par\begingroup\parindent=0pt\leftskip=1cm\rightskip=1cm\parindent=0pt
\baselineskip=11pt
\global\advance\figno by 1
\midinsert
\epsfxsize=#3
\centerline{\epsfbox{#2}}
\vskip -21pt
{\bf Fig.\ \the\figno: } #1\par
\endinsert\endgroup\par
}
\def\figlabel#1{\xdef#1{\the\figno}}
\def\encadremath#1{\vbox{\hrule\hbox{\vrule\kern8pt\vbox{\kern8pt
\hbox{$\displaystyle #1$}\kern8pt}
\kern8pt\vrule}\hrule}}

\def\frac#1#2{{#1 \over #2}}

\def\p{\partial}
\def\semi{\subset\kern-1em\times\;}
\def\bar#1{\overline{#1}}
\def\sqr#1#2{{\vcenter{\vbox{\hrule height.#2pt
\hbox{\vrule width.#2pt height#1pt \kern#1pt \vrule width.#2pt}
\hrule height.#2pt}}}}

\def\p{\partial}

\def\ad{\bar a}

\def\ap{\alpha'}

\def\rt{{\tilde{r}}}

\def\psib{\overline{\psi}}

\def\p{\partial}

\def\pslash{p\!\!\! /}
\def\kslash{k\!\!\! /}
\def\pslash{\partial\!\!\! /}
\def\sh{\hat{\sigma}}

\def\bh{\hat{\beta}}
\def\Aslash{A\!\!\! /}

%
\def\ap{\alpha'}

%



%
\lref\prange{Richard E. Prange and Steven M. Girvin, eds., ``The Quantum Hall
  Effect" (2nd ed.)
(Springer-Verlag, 1990) }
\lref\das{Sankar Das Sarma and Aron Pinczuk, eds., ``Perspectives in Quantum Hall Effects" (John Wiley and Sons, 1997)}
\lref\girvin{
Steven M. Girvin, ``The Quantum Hall Effect: Novel Excitations and Broken Symmetries", arXiv:cond-mat/9907002
}
\lref\karlhede{A. Karlhede, S.A. Kivelson, S.L. Sondhi, ``The Quantum Hall effect: The Article'' in {\it Correlated Electron Systems (Jerusalem Winter School in Theoretical Physics, vol. 9)}, V.J. Emery ed.   (World Scientific, 1992)}

\lref\Shankar{
G.~Murthy and R.~Shankar,
``Hamiltonian theories of the fractional quantum Hall effect",
Rev. Mod. Phys, {\bf 75}, 1101 (2003)
}

\lref\sondhi{S.L. Sondhi, S.M. Girvin, J.P. Carini, D. Shahar, ``Continuous Quantum Phase Transitions'', arXiv:cond-mat/9609279}

\lref\huck{B. Huckestein, ``Scaling Theory of the Integer Quantum Hall Effect", Rev. Mod. Phys., {\bf 67}, 357, (1995)}


\lref\ludwig{A.W.W. Ludwig, M.P.A. Fisher, R. Shankar, G. Grinstein, ``Integer quantum Hall transition: An alternative approach and exact results''  Phys.\ Rev.\  B {\bf 50}, 7526 (1994).
}

\lref\ZhangWY{
  S.~C.~Zhang, T.~H.~Hansson and S.~Kivelson,
  ``An effective field theory model for the fractional quantum hall effect,''
  Phys.\ Rev.\ Lett.\  {\bf 62}, 82 (1988).
}

\lref\ZhangEU{
  S.~C.~Zhang,
  ``The Chern-Simons-Landau-Ginzburg theory of the fractional quantum Hall
  effect,''
  Int.\ J.\ Mod.\ Phys.\  B {\bf 6}, 25 (1992).
}

\lref\sachdev{S. Sachdev, ``Quantum Phase Transitions", Cambridge University Press, (2001) }


\lref\ErdmengerCM{
  J.~Erdmenger, N.~Evans, I.~Kirsch and E.~Threlfall,
  ``Mesons in Gauge/Gravity Duals - A Review,''
  Eur.\ Phys.\ J.\  A {\bf 35}, 81 (2008)
  [arXiv:0711.4467 [hep-th]].
}

\lref\SakaiCN{
  T.~Sakai and S.~Sugimoto,
  ``Low energy hadron physics in holographic QCD,''
  Prog.\ Theor.\ Phys.\  {\bf 113}, 843 (2005)
  [arXiv:hep-th/0412141].
}

\lref\KarchSH{
  A.~Karch and E.~Katz,
  ``Adding flavor to AdS/CFT,''
  JHEP {\bf 0206}, 043 (2002)
  [arXiv:hep-th/0205236].
}

\lref\BabingtonVM{
  J.~Babington, J.~Erdmenger, N.~J.~Evans, Z.~Guralnik and I.~Kirsch,
  ``Chiral symmetry breaking and pions in non-supersymmetric gauge /  gravity
  duals,''
  Phys.\ Rev.\  D {\bf 69}, 066007 (2004)
  [arXiv:hep-th/0306018].
}

\lref\EvansTI{
  N.~Evans, J.~P.~Shock and T.~Waterson,
  ``D7 brane embeddings and chiral symmetry breaking,''
  JHEP {\bf 0503}, 005 (2005)
  [arXiv:hep-th/0502091].
}

\lref\MMT{
  D.~Mateos, R.~C.~Myers and R.~M.~Thomson,
  ``Holographic phase transitions with fundamental matter,''
  Phys.\ Rev.\ Lett.\  {\bf 97}, 091601 (2006)
  [arXiv:hep-th/0605046],
  ``Thermodynamics of the brane,''
  JHEP {\bf 0705}, 067 (2007)
  [arXiv:hep-th/0701132].
}

\lref\KobayashiSB{
  S.~Kobayashi, D.~Mateos, S.~Matsuura, R.~C.~Myers and R.~M.~Thomson,
  ``Holographic phase transitions at finite baryon density,''
  JHEP {\bf 0702}, 016 (2007)
  [arXiv:hep-th/0611099].
}

\lref\AntonyanVW{
  E.~Antonyan, J.~A.~Harvey, S.~Jensen and D.~Kutasov,
  ``NJL and QCD from string theory,''
  arXiv:hep-th/0604017.
}

\lref\AntonyanQY{
  E.~Antonyan, J.~A.~Harvey and D.~Kutasov,
  ``The Gross-Neveu model from string theory,''
  Nucl.\ Phys.\  B {\bf 776}, 93 (2007)
  [arXiv:hep-th/0608149].
}

\lref\DavisKA{
  J.~L.~Davis, M.~Gutperle, P.~Kraus and I.~Sachs,
  ``Stringy NJL and Gross-Neveu models at finite density and temperature,''
  JHEP {\bf 0710}, 049 (2007)
  [arXiv:0708.0589 [hep-th]].
}

\lref\BLL{
  O.~Bergman, G.~Lifschytz and M.~Lippert,
  ``Holographic Nuclear Physics,''
  JHEP {\bf 0711}, 056 (2007)
  [arXiv:0708.0326 [hep-th]],
  ``Magnetic properties of dense holographic QCD,''
  arXiv:0806.0366 [hep-th].
}

\lref\AFJK{
  T.~Albash, V.~G.~Filev, C.~V.~Johnson and A.~Kundu,
  ``A topology-changing phase transition and the dynamics of flavour,''
  Phys.\ Rev.\  D {\bf 77}, 066004 (2008)
  [arXiv:hep-th/0605088],
  ``Global Currents, Phase Transitions, and Chiral Symmetry Breaking in Large
  $N_c$ Gauge Theory,''
  arXiv:hep-th/0605175,
``Finite Temperature Large N Gauge Theory with Quarks in an External Magnetic
  Field,''
  JHEP {\bf 0807}, 080 (2008)
  [arXiv:0709.1547 [hep-th]].
}

\lref\JohnsonVN{
  C.~V.~Johnson and A.~Kundu,
  ``External Fields and Chiral Symmetry Breaking in the Sakai-Sugimoto Model,''
  arXiv:0803.0038 [hep-th].
}

\lref\FilevGB{
  V.~G.~Filev, C.~V.~Johnson, R.~C.~Rashkov and K.~S.~Viswanathan,
  ``Flavoured large N gauge theory in an external magnetic field,''
  JHEP {\bf 0710}, 019 (2007)
  [arXiv:hep-th/0701001].
}

\lref\AharonyDA{
  O.~Aharony, J.~Sonnenschein and S.~Yankielowicz,
  ``A holographic model of deconfinement and chiral symmetry restoration,''
  Annals Phys.\  {\bf 322}, 1420 (2007)
  [arXiv:hep-th/0604161].
}

\lref\PS{
  A.~Parnachev and D.~A.~Sahakyan,
  ``Chiral phase transition from string theory,''
  Phys.\ Rev.\ Lett.\  {\bf 97}, 111601 (2006)
  [arXiv:hep-th/0604173],
``Photoemission with chemical potential from QCD gravity dual,''
  Nucl.\ Phys.\  B {\bf 768}, 177 (2007)
  [arXiv:hep-th/0610247].
}

\lref\ParnachevBC{
  A.~Parnachev,
  ``Holographic QCD with Isospin Chemical Potential,''
  JHEP {\bf 0802}, 062 (2008)
  [arXiv:0708.3170 [hep-th]].
}

\lref\ErdmengerBN{
  J.~Erdmenger, R.~Meyer and J.~P.~Shock,
  ``AdS/CFT with Flavour in Electric and Magnetic Kalb-Ramond Fields,''
  JHEP {\bf 0712}, 091 (2007)
  [arXiv:0709.1551 [hep-th]].
}


\lref\MiranskyA{V.P. Gusyin, V.A. Miransky, S.G. Sharapov, I.A. Shovkovy, ``Excitonic gap, phase transition, and quantum Hall effect in graphene'', Phys. Rev. B {\bf 74}, 195429 (2006), arXiv:cond-mat/0605348}

\lref\MiranskyB{E.V. Gorbar, V.P. Gusynin, V.A. Miransky, ``Toward theory of quantum Hall effect in graphene'', arXiv:0710.3527[cond-mat.mes-hall]}

\lref\GorbarHU{
  E.~V.~Gorbar, V.~P.~Gusynin, V.~A.~Miransky and I.~A.~Shovkovy,
  ``Dynamics in the quantum Hall effect and the phase diagram of graphene,''
  arXiv:0806.0846 [cond-mat.mes-hall].
}

\lref\Checkelsky{J.G. Checkelsky, L. Li, N.P. Ong, ``Divergent resistance at the Dirac Point in graphene: evidence for a transition in high magnetic field'', arXiv:0808.0906[cond-mat.mes-hall]}


\lref\HartnollVX{
  S.~A.~Hartnoll, C.~P.~Herzog and G.~T.~Horowitz,
  ``Building an AdS/CFT superconductor,''
  arXiv:0803.3295 [hep-th].
}

\lref\GubserPX{
  S.~S.~Gubser,
  ``Breaking an Abelian gauge symmetry near a black hole horizon,''
  arXiv:0801.2977 [hep-th].
}

\lref\GubserZU{
  S.~S.~Gubser,
  ``Colorful horizons with charge in anti-de Sitter space,''
  arXiv:0803.3483 [hep-th].
}

\lref\NakanoXC{
  E.~Nakano and W.~Y.~Wen,
  ``Critical magnetic field in AdS/CFT superconductor,''
  arXiv:0804.3180 [hep-th].
}

\lref\AlbashEH{
  T.~Albash and C.~V.~Johnson,
  ``A Holographic Superconductor in an External Magnetic Field,''
  arXiv:0804.3466 [hep-th].
}

\lref\WenPB{
  W.~Y.~Wen,
  ``Inhomogeneous magnetic field in AdS/CFT superconductor,''
  arXiv:0805.1550 [hep-th].
}

\lref\GubserWV{
  S.~S.~Gubser and S.~S.~Pufu,
  ``The gravity dual of a p-wave superconductor,''
  arXiv:0805.2960 [hep-th].
}

\lref\RobertsNS{
  M.~M.~Roberts and S.~A.~Hartnoll,
  ``Pseudogap and time reversal breaking in a holographic superconductor,''
  JHEP {\bf 0808}, 035 (2008)
  [arXiv:0805.3898 [hep-th]].
}


\lref\KarchPD{
  A.~Karch and A.~O'Bannon,
  ``Metallic AdS/CFT,''
  JHEP {\bf 0709}, 024 (2007)
  [arXiv:0705.3870 [hep-th]].
}

\lref\OBannonIN{
  A.~O'Bannon,
  ``Hall Conductivity of Flavor Fields from AdS/CFT,''
  Phys.\ Rev.\  D {\bf 76}, 086007 (2007)
  [arXiv:0708.1994 [hep-th]].
}

\lref\SonVK{
  D.~T.~Son and A.~O.~Starinets,
  ``Viscosity, Black Holes, and Quantum Field Theory,''
  Ann.\ Rev.\ Nucl.\ Part.\ Sci.\  {\bf 57}, 95 (2007)
  [arXiv:0704.0240 [hep-th]].
}

\lref\BergmanSG{
  O.~Bergman, G.~Lifschytz and M.~Lippert,
  ``Response of Holographic QCD to Electric and Magnetic Fields,''
  JHEP {\bf 0805}, 007 (2008)
  [arXiv:0802.3720 [hep-th]].
}

\lref\AlbashBQ{
  T.~Albash, V.~G.~Filev, C.~V.~Johnson and A.~Kundu,
  ``Quarks in an External Electric Field in Finite Temperature Large N Gauge
  Theory,''
  arXiv:0709.1554 [hep-th].
}


\lref\SonYE{
  D.~T.~Son,
  ``Toward an AdS/cold atoms correspondence: a geometric realization of the
  Schroedinger symmetry,''
  Phys.\ Rev.\  D {\bf 78}, 046003 (2008)
  [arXiv:0804.3972 [hep-th]].
}

\lref\BalasubramanianDM{
  K.~Balasubramanian and J.~McGreevy,
  ``Gravity duals for non-relativistic CFTs,''
  Phys.\ Rev.\ Lett.\  {\bf 101}, 061601 (2008)
  [arXiv:0804.4053 [hep-th]].
}

\lref\HerzogWG{
  C.~P.~Herzog, M.~Rangamani and S.~F.~Ross,
  ``Heating up Galilean holography,''
  arXiv:0807.1099 [hep-th].
}

\lref\MaldacenaWH{
  J.~Maldacena, D.~Martelli and Y.~Tachikawa,
  ``Comments on string theory backgrounds with non-relativistic conformal
  symmetry,''
  arXiv:0807.1100 [hep-th].
}

\lref\AdamsWT{
  A.~Adams, K.~Balasubramanian and J.~McGreevy,
  ``Hot Spacetimes for Cold Atoms,''
  arXiv:0807.1111 [hep-th].
}

\lref\schafer{T. Sch\"afer, ``From Equilibrium to Transport Properties of Strongly Correlated Fermi Liquids'', arXiv:0808.0734v1 [nucl-th]}


\lref\BrodieYZ{
  J.~H.~Brodie, L.~Susskind and N.~Toumbas,
  ``How Bob Laughlin tamed the giant graviton from Taub-NUT space,''
  JHEP {\bf 0102}, 003 (2001)
  [arXiv:hep-th/0010105].
}

\lref\BenaCS{
  I.~Bena and A.~Nudelman,
  ``On the stability of the quantum Hall soliton,''
  JHEP {\bf 0012}, 017 (2000)
  [arXiv:hep-th/0011155].
}

\lref\GubserDZ{
  S.~S.~Gubser and M.~Rangamani,
  ``D-brane dynamics and the quantum Hall effect,''
  JHEP {\bf 0105}, 041 (2001)
  [arXiv:hep-th/0012155].
}

\lref\SusskindFB{
  L.~Susskind,
  ``The quantum Hall fluid and non-commutative Chern Simons theory,''
  arXiv:hep-th/0101029.
}

\lref\BergmanQG{
  O.~Bergman, Y.~Okawa and J.~H.~Brodie,
  ``The stringy quantum Hall fluid,''
  JHEP {\bf 0111}, 019 (2001)
  [arXiv:hep-th/0107178].
}

\lref\HellermanYV{
  S.~Hellerman and L.~Susskind,
  ``Realizing the quantum Hall system in string theory,''
  arXiv:hep-th/0107200.
}

\lref\FreivogelVC{
  B.~Freivogel, L.~Susskind and N.~Toumbas,
  ``A two fluid description of the quantum Hall soliton,''
  arXiv:hep-th/0108076.
}

\lref\KeskiVakkuriEB{
  E.~Keski-Vakkuri and P.~Kraus,
  ``Quantum Hall Effect in AdS/CFT,''
  arXiv:0805.4643 [hep-th].
}

\lref\Rey{S.J. Rey, Talk given at {\it Strings 2007} in Madrid (slides available online at\break http://www.ift.uam.es/strings07/040\_scientific07\_contents/transparences/rey.pdf). }


\lref\HartnollAI{
  S.~A.~Hartnoll and P.~Kovtun,
  ``Hall conductivity from dyonic black holes,''
  Phys.\ Rev.\  D {\bf 76}, 066001 (2007)
  [arXiv:0704.1160 [hep-th]].
}

\lref\HartnollIH{
  S.~A.~Hartnoll, P.~K.~Kovtun, M.~Muller and S.~Sachdev,
  ``Theory of the Nernst effect near quantum phase transitions in condensed
  matter, and in dyonic black holes,''
  Phys.\ Rev.\  B {\bf 76}, 144502 (2007)
  [arXiv:0706.3215 [cond-mat.str-el]].
}

\lref\EvansZS{
  N.~Evans and E.~Threlfall,
  ``R-Charge Chemical Potential in a 2+1 Dimensional System,''
  arXiv:0807.3679 [hep-th].
}

\lref\KulaxiziKV{
  M.~Kulaxizi and A.~Parnachev,
  ``Comments on Fermi Liquid from Holography,''
  arXiv:0808.3953 [hep-th].
}

\lref\KachruYH{
  S.~Kachru, X.~Liu and M.~Mulligan,
  ``Gravity Duals of Lifshitz-like Fixed Points,''
  arXiv:0808.1725 [hep-th].
}

\lref\MinicXA{
  D.~Minic and M.~Pleimling,
  ``Non-relativistic AdS/CFT and Aging/Gravity Duality,''
  arXiv:0807.3665 [cond-mat.stat-mech].
}

\lref\KarchFA{
  A.~Karch, D.~T.~Son and A.~O.~Starinets,
  ``Zero Sound from Holography,''
  arXiv:0806.3796 [hep-th].
}

\lref\JejjalaJY{
  V.~Jejjala, D.~Minic, Y.~J.~Ng and C.~H.~Tze,
  ``Turbulence and Holography,''
  arXiv:0806.0030 [hep-th].
}

\lref\BhattacharyyaJI{
  S.~Bhattacharyya, R.~Loganayagam, S.~Minwalla, S.~Nampuri, S.~P.~Trivedi and S.~R.~Wadia,
  ``Forced Fluid Dynamics from Gravity,''
  arXiv:0806.0006 [hep-th].
}

\lref\BhattacharyyaXC{
  S.~Bhattacharyya {\it et al.},
  ``Local Fluid Dynamical Entropy from Gravity,''
  JHEP {\bf 0806}, 055 (2008)
  [arXiv:0803.2526 [hep-th]].
}

\lref\BhattacharyyaJC{
  S.~Bhattacharyya, V.~E.~Hubeny, S.~Minwalla and M.~Rangamani,
  ``Nonlinear Fluid Dynamics from Gravity,''
  JHEP {\bf 0802}, 045 (2008)
  [arXiv:0712.2456 [hep-th]].
}


\lref\RosensteinNM{
  B.~Rosenstein, B.~Warr and S.~H.~Park,
  ``Dynamical symmetry breaking in four Fermi interaction models,''
  Phys.\ Rept.\  {\bf 205}, 59 (1991).
}

\lref\BreitenlohnerBM{
  P.~Breitenlohner and D.~Z.~Freedman,
  ``Positive Energy In Anti-De Sitter Backgrounds And Gauged Extended
  Supergravity,''
  Phys.\ Lett.\  B {\bf 115}, 197 (1982),
  ``Stability In Gauged Extended Supergravity,''
  Annals Phys.\  {\bf 144}, 249 (1982).
}

\lref\GrossJV{
  D.~J.~Gross and A.~Neveu,
  ``Dynamical Symmetry Breaking In Asymptotically Free Field Theories,''
  Phys.\ Rev.\  D {\bf 10}, 3235 (1974).
}

\lref\DunneQY{
  G.~V.~Dunne,
  ``Aspects of Chern-Simons theory,''
  arXiv:hep-th/9902115.
}

\lref\parity{
  A.~J.~Niemi and G.~W.~Semenoff,
  ``Axial Anomaly Induced Fermion Fractionization And Effective Gauge Theory
  Actions In Odd Dimensional Space-Times,''
  Phys.\ Rev.\ Lett.\  {\bf 51}, 2077 (1983).
  A.~N.~Redlich,
  ``Gauge Noninvariance And Parity Nonconservation Of Three-Dimensional
  Fermions,''
  Phys.\ Rev.\ Lett.\  {\bf 52}, 18 (1984),
``Parity Violation And Gauge Noninvariance Of The Effective Gauge Field
  Action In Three-Dimensions,''
  Phys.\ Rev.\  D {\bf 29}, 2366 (1984).
}

\lref\LifschytzIQ{
  G.~Lifschytz,
  ``Comparing D-branes to Black-branes,''
  Phys.\ Lett.\  B {\bf 388}, 720 (1996)
  [arXiv:hep-th/9604156].
}


\lref\Weicurrent{H.P. Wei, L.W. Engel, and D.C. Tsui, 1994, Phys. Rev. B {\bf 50}, 14609}

\lref\Weitemp{H.P. Wei, D.C. Tsui, M.A. Paalanen, and A.M.M. Pruisken, 1988, Phys. Rev. Lett. {\bf 61},
1294}

\lref\Wong{L.W. Wong, H.W. Jiang, N. Trivedi, and E. Palm, 1995, Phys. Rev. B {\bf 51}, 18033}

\lref\Engel{L.W. Engel, D. Shahar, C. Kurdak, and D.C. Tsui, 1993, Phys. Rev. Lett. {\bf 71}, 2638}


\Title{\vbox{\baselineskip12pt
}} {\vbox{\centerline {Gravity Dual of a Quantum Hall Plateau Transition}}}
\centerline{
Joshua L. Davis$^{\dagger}$\foot{davis@physics.ucla.edu}, Per
Kraus$^{\dagger}$\foot{pkraus@ucla.edu} and Akhil Shah$^\dagger$\foot{akhil@physics.ucla.edu}}
\bigskip
\centerline{${}^\dagger$\it{Department of Physics and
Astronomy, UCLA,}}\centerline{\it{ Los Angeles, CA 90095-1547,
USA.}}

\baselineskip15pt

\vskip .3in

\centerline{\bf Abstract}

We show how to model the transition between distinct quantum Hall plateaus in terms of D-branes
in string theory.  A low energy theory of $2+1$ dimensional fermions is obtained by
considering the D3-D7 system, and the plateau  transition corresponds to moving the branes
through one another.  We study the transition at strong coupling using gauge/gravity duality
and the probe approximation.  Strong coupling leads to a novel kind of plateau transition:
at low  temperatures  the transition remains discontinuous due to the effects of dynamical symmetry breaking and mass generation, and at high temperatures  is only partially smoothed out.

\Date{September, 2008}
\baselineskip14pt

\newsec{Introduction}

The quantum Hall effect (QHE) is one of the most fascinating phenomena in condensed matter physics, and
has commanded the attention of theorists and experimentalists alike for nearly 30 years.\foot{For  reviews and references to the original literature see {\it e.g.}, \refs{\prange,\das,\girvin,\Shankar,\karlhede}.}  It is also
a very general effect, relying on just a few underlying properties:  any system
of effectively $2+1$ dimensional charged particles with broken parity symmetry and  an energy gap for charged excitations can
be expected to exhibit quantum Hall behavior.   In the condensed matter setting, these properties typically are
realized by electrons in semiconductor junctions subject to low temperature and large magnetic fields.   The magnetic
field breaks the parity symmetry, and the energy gap arises from a combination of Landau level quantization,
disorder, and (at least in the fractional case) electron-electron interactions.    These properties also
arise rather naturally in terms of D-brane configurations in string theory, which is our interest here.

An especially interesting aspect of quantum Hall physics, and one which is still not completely understood
to this day, is the transition between distinct plateaus.   At zero temperature, as we dial some
control parameter such as the magnetic field, the Hall conductivity $\sigma_{xy}$ is seen to jump from
one quantized value to another.   The longitudinal conductivity vanishes except precisely at the
transition point. The transition point is realized when the energy gap of the system vanishes, and
the theory at the transition is a quantum critical point, meaning that it can be described by a scale
invariant theory in $2+1$ dimensions \sachdev\ (to be distinguished  from a critical point controlled by thermal fluctuations,  which is described by a scale invariant Euclidean theory living in the spatial dimensions).
\fig{Schematic illustration of quantum Hall plateau transitions at small but finite
temperature.   The longitudinal conductivity increases sharply at the transition.
The slope of the transverse conductivity defines the critical exponent $\nu z$.  }{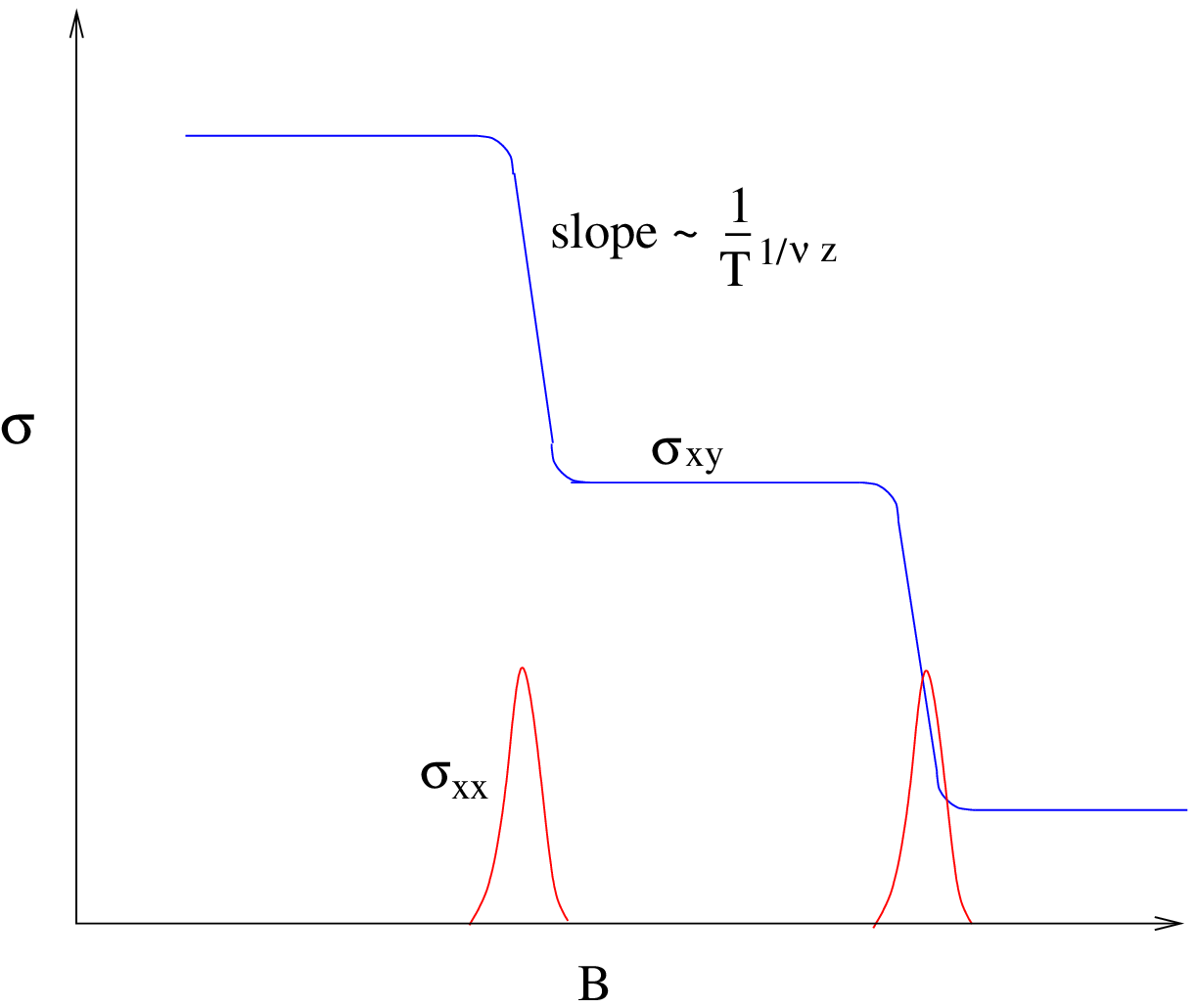}{3.0truein}

The transition is smoothed out at finite temperature.  The slope of $\sigma_{xy}$ versus $B$ becomes
finite at the transition, obeying a scaling behavior
\eqn\ga{ {\p \sigma_{xy} \over \p B} \sim {1\over T^{1/ \nu z}}~.}
Here $\nu$ parameterizes the divergence of the spatial correlation length and $z$ is the dynamical critical exponent, determining the relative scalings of time and space
at the critical point.  The exponents $\nu$ and $z$ have been measured in various integer and fractional plateau
transitions, yielding $z \approx 1$ and   $\nu \approx 2.4$.     Since the same values are measured
for different transitions in different materials, there is evidence of a universal critical theory. See the review  \sondhi\  for further discussion of the measurement of these scaling exponents and their relation to  quantum criticality.

On the theoretical side, the transition is usually modeled as a disorder mediated localization-delocalization transition;  see the article by Pruisken in \prange\ and the review  \huck.   The essential idea is that the random potential due to impurities modifies the Landau level spectrum, localizing all states except for a finite number which remain delocalized.  The plateau transition
occurs as the Fermi level crosses the energy of a delocalized state, resulting in a vanishing energy gap and a sharp change in the conductance.
The result $\nu \approx 2.4$ can be computed numerically given a specific model of the disorder.   These
computations are usually carried out for free electrons; the study of disorder in interacting electron
systems, relevant for the fractional QHE, is much more difficult and hence less understood.

Quantum Hall plateau transitions also occur in idealized systems without disorder.   Perhaps the simplest
such example is that of free relativistic fermions in $2+1$ dimensions \ludwig.   Here the control
parameter is the fermion mass $m$.  Even without a background magnetic field or charge density,
the  conductivity is nonzero due to vacuum polarization, with the transverse conductivity
exhibiting half-integer quantization.  A fermion mass term in $2+1$ dimensions is parity odd, and a plateau
transition is realized by smoothly changing the sign of the mass.  The quantum critical point in this
case is just the theory of a free massless fermion, with critical exponents $z=\nu=1$.   In \ludwig\
this model was used as a starting point to which disorder could be added.

A low energy theory of free fermions  in $2+1$ dimensions is easily obtained in string theory
by considering a $2+1$ dimensional intersection of D3-branes and D7-branes at zero string coupling.
This is a  $\# ND=6$ system, which is  non-supersymmetric but tachyon free.  The fermion mass is equal
to the separation of the branes in the one common transverse direction.  The T-dual D4-D8 system
has been heavily studied in recent years as a model of QCD with chiral fermions (the Sakai-Sugimoto
model \SakaiCN), and the D3-D7 system can be studied using similar methods.  More recently, the D3-D7 system
was proposed by Rey to give a description of the quantum critical point of graphene \Rey\ (and article to appear).   As will be
discussed, we will not actually find a quantum critical point in our setup (both on the gravity side and in our field theory model at strong coupling), but our methods and some of our
results overlap with the earlier work of Rey.

In D-brane terms, the plateau transition is realized by starting with a nonzero transverse separation, taking it to zero, and then reversing its sign.\foot{In other words, passing the $D7$-brane and $D3$-brane stacks through each other, with the critical point occurring when they are coincident.}  We will be interested in the dual gravity description of this
operation, valid in the strong coupling regime.  In particular, in the limit of large  $N_{D3}$ and
$g_s N_{D3}$,  we can replace the D3-branes by their corresponding supergravity solution.  For
$N_{D7} \ll N_{D3}$ we can treat the D7-branes as probes in this background geometry \KarchSH.   There is an extensive
literature on using this probe approximation to study QCD with quarks, as well as related theories; see {\it e.g.} \refs{\BabingtonVM,\EvansTI,\AharonyDA,\MMT,\PS,\DavisKA,\ParnachevBC,\BLL,\BergmanSG,\AFJK,\ErdmengerBN,\AlbashBQ,\JohnsonVN,\FilevGB} for studies of the phase structure of QCD-like theories with fundamental matter and \ErdmengerCM\ for a thorough review of the literature. The
nicest feature of this setup is that dynamical symmetry breaking and mass generation have a simple
geometrical manifestation \refs{\SakaiCN,\AntonyanVW,\AntonyanQY}.

We study the plateau transition in the interacting theory, both on the field theory side and in gravity.
The main result is that at strong coupling the nature of the transition is strongly affected by the interactions, as can be seen by comparing Figs. 1 and 6.    In particular, the transition is not completely smoothed out by turning on a nonzero temperature.
The transverse conductivity instead continues to exhibit a finite jump, due to the underlying dynamical mass generation.  On the field theory side we model the interactions in terms of four-Fermi interactions.
Above some critical temperature $T_c$ the dynamically generated mass vanishes and the transition is
completely smoothed out.   The situation on the gravity side is somewhat more complicated.    There is
again a critical temperature $T_c$, but now for $T>T_c$ the transition is only partially smoothed out.
In geometrical terms, the smoothed out region corresponds to a family of probe geometries which fall through
the horizon of the finite temperature black hole geometry.    Branes which fall through black hole horizons
exhibit dissipative behavior, and in particular a nonzero longitudinal conductivity, which we compute
using standard techniques.

It is of course natural to wonder whether the strong coupling effects that we find could have any analog
in the real world.   While we do not know the answer to this, it is worth noting that dynamical symmetry
breaking has been proposed to play a role in the quantum Hall effect in graphene \refs{\MiranskyA,\MiranskyB,\GorbarHU} including the recently observed quantum Hall insulator transition \Checkelsky. It would be
interesting to explore this possibility further.

The study of holographic and string theory models directed at addressing problems in condensed matter systems is a burgeoning field. Recent attention has been directed towards holographic realizations of superconductors \refs{\GubserPX,\HartnollVX,\GubserZU,\NakanoXC,\AlbashEH,\WenPB,\GubserWV,\RobertsNS} and the AdS/Galiliean CFT correspondence approach \refs{\SonYE,\BalasubramanianDM,\HerzogWG,\MaldacenaWH,\AdamsWT} to the study of fermions at unitarity.\foot{See \schafer\ for a brief summary of the unitarity fermion system.} Further topics of holographic investigation have included Fermi liquids \refs{\KarchFA,\KulaxiziKV}, non-Lorentz invariant fixed points \KachruYH, non-equilibrium statistical mechanics \MinicXA, and non-linear hydrodynamics \refs{\BhattacharyyaJC,\BhattacharyyaXC,\BhattacharyyaJI,\JejjalaJY}. There are also a number of previous works discussing the QHE.\foot{The classical hall conductivity was studied using AdS/CFT in \refs{\HartnollAI,\HartnollIH}.} Older papers, not  involving the AdS/CFT correspondence, include \refs{\BrodieYZ,\BenaCS,\GubserDZ,\SusskindFB,\BergmanQG,\HellermanYV,\FreivogelVC} while the recent \KeskiVakkuriEB\ provides an effective holographic model of the Zhang-Hansson-Kivelson theory \refs{\ZhangWY,\ZhangEU} of  fractional quantum Hall plateaus.

The remainder of the paper is organized as follows. Section 2 is devoted to field theory models of the plateau transition. The free fermion theory and four-fermion interactions are discussed, at zero and finite temperatures, with the role of the parity anomaly and induced Chern-Simons terms emphasized. In Section 3, we introduce our model of intersecting D3 and D7 branes and describe probe solutions of the D7 branes in the D3 brane background. The dual geometric realization of the plateau transition is discussed. The longitudinal and Hall conductivities of our scenario are computed in Section 4. Finally, we close with some comments in Section 5.

\newsec{Field theory models}

\subsec{Free fermions}

In this section we collect some facts about free fermions in $2+1$ dimensions (see also \refs{\DunneQY,\RosensteinNM}).    The basic
building block is a two-component Majorana spinor, but for present purposes it is more convenient to combine these into two-component complex spinors.   The Dirac action for $N$ such fields,
\eqn\aa{ {\cal L} = \psib^i i \pslash \psi_i }
is invariant under a $O(2N)$ global symmetry,  as well as $P$, $C$ and $T$.  Parity flips one
of the spatial coordinates, say $x^1$, acting on the spinors as $\psi_i \rightarrow \gamma^1 \psi_i$, and leaves \aa\ invariant.   However, an important point is that the  mass term $m\psib \psi$ is parity {\it odd}.

Although there is no chiral symmetry in $2+1$ dimensions, a somewhat analogous symmetry is realized by considering an action of the form
\eqn\ab{ {\cal L} =  \psib^i_{+} i \pslash \psi_{+i} +\psib^i_{-} i \pslash \psi_{-i} +
m \psib^i_{+}\psi_{+i}- m \psib^i_{-}\psi_{-i}~.}
This action is parity even (taking parity to swap the $\pm$ labels) and has a $O(2N)_+ \times
O(2N)_-$ global symmetry, enhanced to $O(4N)$ for $m=0$.

In \ab\ we are labeling  massive fermions as $\pm$ according to the sign of their mass
terms.  For massless fermions there is no such distinction in the classical action, but
one arises in the quantum theory due to the parity anomaly \parity, as we'll see  in more detail below.  For the moment, we note that this arises from regulating UV divergences in one-loop diagrams in the presence of an external gauge field.  Pauli-Villars provides a gauge
invariant regularization, but breaks parity due to the need to choose a sign for the mass
of the regulator fermion.  The anomaly only afflicts theories with an odd number of fermions,
since with an even number we can preserve parity by choosing equal numbers of positive and negative mass regulator fields.

We now couple the fermions to a $U(1)$ gauge field (dropping the $\pm$ label),
\eqn\ac{ {\cal L} = \psib^i (i \pslash-m-\Aslash) \psi_i }
and compute the one-loop effective action for the gauge field,
\eqn\ad{S_{eff} = \int\! {d^3p \over (2\pi)^3 }\Pi^{\mu\nu}(p)A_\mu(p)A_\nu(-p)~.}
The  vacuum polarization diagram has parity even and odd parts:
\eqn\ae{\eqalign{ \Pi^{\mu\nu}(p) &= \int\! {d^3 k \over (2\pi)^3} {\rm Tr}\left[ \gamma^\mu
{p\!\!\!/ + \kslash - m\over (p+k)^2-m^2} \gamma^\nu{\kslash -m \over k^2-m^2} \right] \cr
&=  (p^\mu p^\nu -g^{\mu\nu}p^2) \Pi_{\rm even}(p) + \epsilon^{\mu\nu\lambda}p_\lambda \Pi_{\rm odd}(p)~.}}
A standard computation in dimensional regularization yields
\eqn\af{\eqalign{ \Pi_{\rm even}(p)& =-{N\over 2\pi}  {1 \over \sqrt{p^2}} \left[ {1\over 2} {|m|\over \sqrt{p^2}} - ({m^2 \over p^2} -{1\over 4})\arctan \left(\sqrt{p^2} \over 2 |m|\right)  \right]  \cr \Pi_{\rm odd}(p)& =-{N \over 2\pi} {m \over \sqrt{p^2} } \arctan\left(\sqrt{p^2} \over 2|m| \right)~.    }}

The small momentum limit of $\Pi_{\rm odd}$ translates into a Chern-Simons term for
the gauge field,
\eqn\ag{ S_{CS} = {k\over 4\pi} \int\! d^3 x~ \epsilon^{\mu\nu\lambda}A_\mu \p_\nu A_\lambda}
with
\eqn\ah{ k=\cases{
 {N\over 2} {\rm sgn}(m) & $m\neq 0$ \cr
 0 & $m=0$
 }}
%
Equivalently, this yields a transverse
DC conductivity $\sigma^T_{ij} = {k\over 2\pi}\epsilon_{ij}$.

The parity anomaly follows from this result.  When we compute the effective action
for a massless fermion together with its Pauli-Villars regulator field, we will
obtain a parity violating Chern-Simons term of coefficient $k= \pm {1\over 2}$ \parity.
Parity can be restored by adding by hand a Chern-Simons term of opposite sign, but
this violates invariance under  gauge transformations with support at infinity.
The parity anomaly is the statement that we cannot preserve both parity and gauge
invariance simultaneously.

From $\Pi_{\rm even}(p)$ we read off the longitudinal AC conductivity from the Kubo
formula:
\eqn\ai{ \sigma^L_{ij}(\omega) = {\rm Im} {1 \over \omega} \Pi^{ij}_{\rm even}(p) ~,\quad p^\mu = (\omega,0,0)~.}
In particular, the DC conductivity is
\eqn\aj{\sigma^L_{ij}(0)=\cases{
0 & $m\neq 0$ \cr
 {N\over 16}\delta{ij}& $m=0$
 }}
%

The computation of the Chern-Simons term can be generalized to finite temperature, with the result \DunneQY\
\eqn\ak{k(T) = N\tanh \left(|m|\over 2T\right){\rm sgn}(m)~.}

\subsec{Quantum Hall plateau transition}

This simple system of free fermions  exhibits a quantum Hall plateau transition due
to the discontinuous behavior of the conductivities as a function of mass \ludwig.    Suppose
we start with $N$ fermions, all with positive mass.   Now evolve one of the masses,
say $m_1$, smoothly to a negative value.   The conductivities as a function of $m_1$
are
\eqn\al{\eqalign{\sigma_{xx}&=\cases{0 & ~~~$m_1>0$\cr {1\over 16} & ~~~$m_1= 0$ \cr 0 & ~~~$m_1<0$}\cr
\sigma_{xy}&=\cases{{N\over 4\pi} & $m_1>0$\cr {N-1\over 4\pi} & $m_1= 0$ \cr {N-2\over 4\pi} & $m_1<0$}}}
%
%
The jump $\Delta \sigma_{xy} = {1 \over 2\pi} = {e^2 \over h}$  is the same as observed
in the experimentally realized integer quantum Hall effect.    At $m_1=0$ we have
a quantum critical point, which in this case is just the trivial theory of a free
massless fermion.   Since this is a Lorentz invariant theory, it has a corresponding
dynamic critical exponent $z=1$, which turns out to be the value observed experimentally (see footnote at end of next paragraph).

At zero temperature the jump in conductivities occurs precisely at $m_1=0$.  Finite
temperature smooths out the transition, giving it a width (from \ak)  $\Delta m_1 \sim T$, as indeed follows from dimensional analysis.   This behavior defines a critical exponent $\nu$ in terms of the derivative of the conductivity at the transition point,
\eqn\am{ {\p \sigma_{xy} \over \p {m_1} } \Big|_{m_1=0}  \sim {1\over T^{1/\nu z}}~. }
Since $z=1$, we find  $\nu=1$ for free fermions.   By contrast, the experimentally measured value of $\nu$ for a wide class of quantum Hall transitions is $\nu \approx 2.4$.\foot{Strictly speaking, independent experiments measure the combinations of exponents $1/\nu z \approx .42$ \refs{\Weitemp,\Wong,\Engel}  and $\nu (z+1)\approx 4.6$ \Weicurrent.}

\subsec{Including interactions: the Gross-Neveu model in $D=2+1$}

Our gravitational description of the plateau transition will be dual to that in a
strongly coupled field theory.  With this in mind, we now consider including
interactions in our field theory model.

We consider the following Lagrangian \GrossJV,
\eqn\ba{\eqalign{{\cal L} &= i \psib^i \pslash \psi_i +{g^2 \over 2N} (\psib^i \psi_i)^2 \cr & \cong i \psib^i \pslash \psi_i -\sigma \psib^i \psi_i -{N\sigma^2 \over 2g^2}~. }}
In the second line we have written an equivalent Lagrangian by  introducing the auxiliary field $\sigma$.  Integrating out $\sigma$ gives back the original Lagrangian in the first  line.   We are taking the fermions to be two-component complex spinors.

%
%

\lref\GrossVU{
  D.~J.~Gross,
  ``Applications Of The Renormalization Group To High-Energy Physics,''
{\it  In *Les Houches 1975, Proceedings, Methods In Field Theory*, Amsterdam 1976, 141-250}
}

We will take $N\gg 1$ and work to leading order in the $1/N$ expansion.
This theory is nonrenormalizable in weak coupling perturbation theory around $g=0$,
since $(\psib \psi)^2$ has mass dimension $4$.    On the other hand, it has long been known that this theory has a nontrivial UV fixed point \refs{\GrossVU,\RosensteinNM}.  We will recall how this appears below, although we will not make significant use of this fact, since there is no
obvious reason why the interacting fermion models realized in our D-brane construction
lie near this fixed point.

The vacuum structure of the theory can be studied in terms of the  effective  potential for $\sigma$,  obtained by integrating out the fermions:
\eqn\usb{{1\over N}V_{eff}(\sigma) = {\sigma^2 \over 2g^2} + \tr \ln(i\pslash -\sigma)~. }
It is easiest to first compute the derivative
\eqn\bc{\eqalign{  {1\over N}V_{eff}'(\sigma) &={\sigma \over g^2} -2\sigma \int_{|p_E|<\Lambda} \! {d^3 p_E \over (2\pi)^3} {1\over p_E^2 +\sigma^2}\cr &={\sigma \over g^2} -{ \sigma
\over \pi^2}\left[\Lambda - \sigma \arctan \left({\Lambda \over \sigma}\right)\right] } }
where we have Wick rotated the integral and imposed a hard UV cutoff.

In the large $N$ limit we identify vacua by solving $V'_{eff}=0$ using the
effective potential written above.   Corrections to $V_{eff}$ due to $\sigma$
fluctuations are suppressed since the $\sigma$ propagator carries a factor of $1/N$.
Since $\langle \sigma \rangle = -{2g^2 \over N} \langle \psib_i \psi_i\rangle$, parity is spontaneously broken for $\langle \sigma\rangle \ne 0$.

There is always a solution of $V_{eff}'=0$ at $\sigma=0$, but it is only stable
for sufficiently small $g^2\Lambda $.    Since  ${1\over N}V_{eff}''(0)= {1\over g^2}-{\Lambda\over \pi^2}$, we see that the   critical coupling is
\eqn\bd{ g_{c}^2\Lambda  = {\pi^2}~. }
For $g<g_c$ the symmetry preserving vacuum is stable, while for $g>g_c$ it is unstable.

Nontrivial solutions of $V'_{eff}=0$  obey the following  gap equation
\eqn\bez{ {\sigma \over \Lambda} \arctan\left({\Lambda \over \sigma}\right) = 1- {\pi^2\over g^2 \Lambda}~.}
The function on the left always lies between $0$ and $1$, and so a solution only exists
for $g>g_c$.

Combining the above results we find the phase structure.  For $g<g_c$ we have
a stable phase of unbroken symmetry.  For $g>g_c$ we have a broken symmetry phase,
which is readily seen to be stable.  The  vev $\langle \psib^i \psi_j \rangle = v\delta^i_j$ breaks parity.      As $g\rightarrow g_c$ from above, the magnitude of the condensate
smoothly slides over to zero, corresponding to a second order phase transition (quantum critical point).

The theory near the critical point takes a simple form.   Define a mass scale $M$
by solving \bez\ near the critical coupling:  $M = 2\pi \left({1 \over g_c^2 \Lambda}-{1\over g^2 \Lambda}\right)\Lambda$.  The limit $\Lambda \rightarrow \infty$ at fixed $M$ then yields an  effective potential
\eqn\bfz{V_{eff}(\sigma)= {N\over 2\pi} \left(-{1 \over 2}M\sigma^2 +{1\over 3}\sigma^3\right)~.}
By working out the full momentum dependent $\sigma$ propagator one finds that it has
large momentum behavior $1/p$, and so $\sigma$ acquires scaling dimension $1$ at the
UV fixed point, as indicated by \bfz.   This also shows that the  fixed point persists order by order in the $1/N$ expansion \refs{\GrossVU,\RosensteinNM} (the value of $g_c^2 \Lambda$ changes with $N$ of course.)

\subsec{Finite temperature}

Sufficiently large temperature will restore the spontaneously broken parity
symmetry.    This can be studied by extending the gap equation \bez\ to finite temperature.  To motivate its form,  go back to the expression for $V'_{eff}$ in the top line of \bc, and perform the $p_E^0$ integration to get the zero temperature gap equation
\eqn\ca{ {\sigma \over g^2 } -  \sigma\int_{|p_E|<\Lambda} {d^2 p_E \over (2\pi)^2}{1 \over E_p  }=0}
with $E_p = \sqrt{p_E^2 + \sigma^2}$.  Note that $\Lambda$ is now a cutoff on
spatial momentum rather than three momentum.    One can think of this equation as
a self-consistent mass formula, in which the assumed fermion mass $\sigma$ is set
equal to the value obtained from interaction of the fermion with the Dirac sea.
This intuition\foot{which is of course in agreement with the result obtained by a systematic finite temperature field theory computation.} makes it easy to write down the   finite temperature version, which is
\eqn\cb{ {\sigma \over g^2 } - \sigma \int_{|p_E|<\Lambda} {d^2 p_E \over (2\pi)^2}{1 \over E_p}\Big(1 - 2n(\beta)\Big) =0 }
where $n(\beta)$ is the Fermi-Dirac distribution
\eqn\cc{ n(\beta) = {1 \over e^{\beta E_p}+1 }~.}
In terms of the dimensionless variables
\eqn\cd{ \sh = \sigma/\Lambda~,\quad \bh = \Lambda \beta }
the gap equation takes the form
\eqn\ce{ {\sh \over g^2 \Lambda} - {\sh\over 2\pi}\int_0^1 \! dx {x \over
 \sqrt{x^2 +\sh^2}} \left( {e^{\bh \sqrt{x^2 +\sh^2}} -1 \over e^{\bh \sqrt{x^2 +\sh^2}} +1 }   \right) =0~.}

The gap equation admits nontrivial solutions for sufficiently large $g^2 \Lambda$
and $\bh$.   Symmetry is restored at weak coupling or  high temperature in a continuous phase transition.
The phase diagram is shown in the figure.
\fig{Solution to the finite temperature gap equation.  Note that $\sigma$ smoothly goes to zero with increasing $T$ or decreasing $g^2$. }{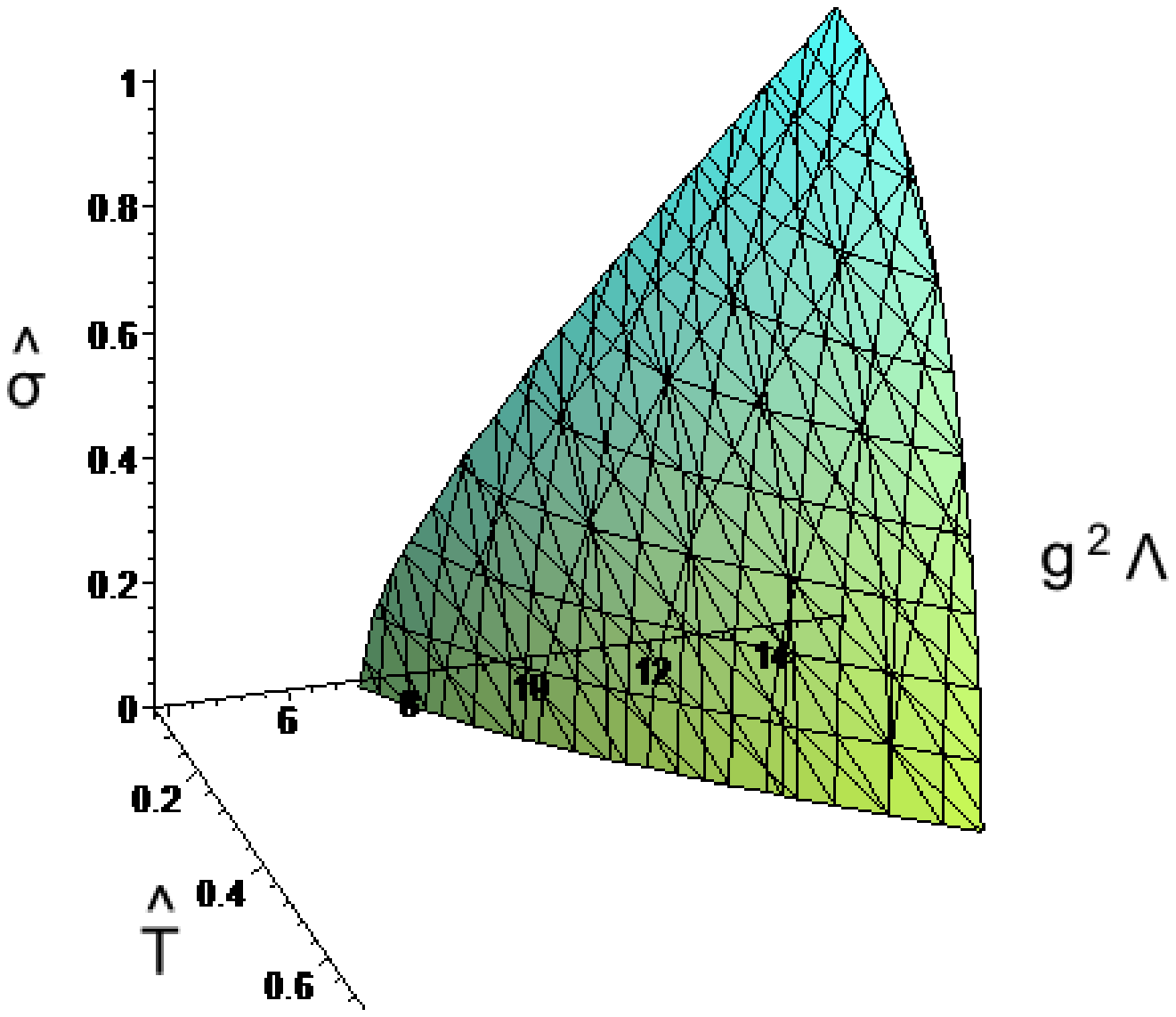}{4.1truein}
\subsec{Plateau transition in the Gross-Neveu model}

Now we describe how the plateau transition proceeds in this interacting theory.
The story is different depending on whether we are in the weak or strong coupling
phase.  The strong coupling case, $g>g_c$, turns out to be the one relevant to
our D-brane construction, so we focus on it.  In this case, at zero temperature
parity is broken and $\langle \sigma \rangle = \pm v$.  The spectrum consists of
$N$ massive fermions, with dynamically generated masses $m_{dyn} \sim \pm v$.
From this it follows that if we couple the theory to a $U(1)$ gauge field, there will be  a Chern-Simons term present with coefficient $k = \pm {N\over 2}$, and the
corresponding transverse conductivity.

 As in the free fermion example, we
now give small bare masses to the fermions.  In our D-brane theory all the fermions
have the same mass $m$, and so we choose that here as well.     We now examine the
effect of smoothly changing the sign of $m$ from positive to negative.

In the low temperature phase in which parity is spontaneously broken, the effect
of $m$ is simply to break the degeneracy between the two vacua related by parity.
So as $m$ passes through zero,  the true vacuum jumps from $\langle \sigma \rangle =v$ to $\langle \sigma \rangle =-v$.    The sign of the Chern-Simons term also jumps, and
so this is a quantum Hall plateau transition.  Unlike in the free fermion example,
the transition remains discontinuous at low temperature due to the dynamical symmetry breaking.

To make the transition continuous we need to turn on a finite temperature.
In particular, for given coupling $g$, there exists a critical temperature $T_c(g)$
at which the broken symmetry is restored and the plateau transition is  continuous.
In analogy with \am\ we now define the critical exponent $\nu z$ as\foot{Since this
is finite temperature transition rather than a quantum phase transition there is no
meaningful dynamic critical exponent $z$, so we only speak of the combination $\nu z$.}
\eqn\cf{ {\p \sigma_{xy} \over \p m}\Big|_{m=0} \sim {1\over (T-T_c)^{1/\nu z} }~.}
To compute this exponent we should compute the induced Chern-Simons term from
the current-current two-point function, as in \ak.  The difference from the
previous computation is that we now have an interacting theory, and so the
one-loop vacuum polarization diagram receives higher loop corrections from
internal $\sigma$ propagators.   Therefore, the exponent $\nu$ will depend on
 $g^2\Lambda$ and $N$ in a nontrivial way.  However, in the large $N$ limit
the higher loop corrections are suppressed by factors of $1/N$, and the computation
reduces to the free case.  We conclude that $\nu z = 1 + O({1\over N})$.

The plateau transition becomes a quantum phase transition in the weak coupling regime,
$g < g_c$, since then $T_c=0$.   Now $z=1$ by Lorentz invariance, and $\nu=1 + O({1\over N})$ as above.

\newsec{Gravity side:  Quantum Hall-ography}

\subsec{Weak coupling brane setup}

We consider $N_7$  D7-branes intersecting $N_3$ D3-branes  over $2+1$ dimensions, according to
\eqn\da{ \matrix{ &0&1&2&3&4&5&6&7&8&9 \cr D3: &\times&\times&\times&\times&&&&&&  \cr
D7: &\times&\times&\times&&\times&\times&\times&\times&\times&&} }
We take the D3-branes to be located at $x^9=0$, and the D7-branes at   $x^9=L$.
This is a non-supersymmetric brane intersection, the number of Neumann-Dirichlet directions being six.   There is no tachyon, since the ground state energy in the NS sector is positive
for $\#ND=6$.    For $L=0$ the massless spectrum consists of $N=N_3 N_7$ complex two-component spinors coming from the Ramond sector.  The NS sector contributes only massive states. This is the setup presented by Rey in \Rey.

We take the  directions $x^3, x^4, \ldots, x^8$ to be noncompact, so the D3 and D7 gauge couplings
vanish when reduced to the $2+1$ dimensional intersection.    We are interested in  the low energy theory of the fermions confined to the intersection.
These fermions have various interaction terms suppressed by powers of the string scale,
{\it e.g.} ${\cal L} \sim  l_{st}^{2n-3}(\psib \psi)^n $.  This theory can be analyzed by the methods of Section 2.3, though the details of course depend on the
precise form of the interactions.

We will be interested in the spontaneous breaking of parity at strong coupling.
We can define parity to be a $\pi$ rotation in the $x^8-x^9$ plane.   This
acts as $D7 \rightarrow  \overline{D7}$ and $L\rightarrow -L$, which matches
the action of parity in the field theory, where $\psi_+ \rightarrow \psi_-$ and
$m\rightarrow -m$, as discussed in Section 2.1.

The gravity description will be studied in the probe approximation \KarchSH, in which
$N_7 \ll N_3$.   In this limit,  for  large $N_3$ and $gN_3$, one can treat the
D7-branes as probes in the supergravity geometry produced by the D3-branes.   Such
a description has been heavily studied in the case of supersymmetric brane intersections, such as D3-branes and D7-branes with a $3+1$ dimensional intersection.
In the non-supersymmetric case the basic idea is the same, though some of the behavior differs.

\subsec{Gravity setup}

The background geometry is  the asymptotically flat, finite temperature D3-brane
solution
\eqn\db{ds^2 = f^{-{1\over 2}}(-h dt^2+ d\vec{x}^2)+f^{1\over 2}(h^{-1}dr^2+r^2 d\Omega_5^2)}
with
\eqn\dc{\eqalign{ f& = 1+{R^4 \over r^4}\cr h &=1-{r_+^4 \over r^4}}}
and we  write
\eqn\dd{ d\Omega_5^2 = d\psi^2 +\sin^2 \psi d\Omega_4^2~.}
To compare with \da\ we note that
\eqn\de{ x^9 = r\cos \psi~.}
The number of D3-branes and the Hawking temperature  are given by
\eqn\df{ N_3 = {R^2 \sqrt{r_+^4+R^4}\over 4\pi g \alpha'^2}~,\quad T = {r_+ \over \pi \sqrt{r_+^4+R^4}}~.}

Our D7-probe spans $x^0, x^1, x^2$, wraps the $S^4$ displayed in \dd, and lies on a
curve $r= r(\psi)$.   The induced worldvolume metric is
\eqn\dg{ ds_8^2 = f^{-{1\over 2}}(-h dt^2+ dx^i dx^i)+f^{1\over 2}\left(h^{-1}r'^2+r^2\right)d\psi^2 +f^{1\over 2}r^2 \sin^2 \psi  d\Omega_4^2~,}
with $i=1,2$.
In the absence of worldvolume gauge fields, the worldvolume action is given by the  Born-Infeld action as
\eqn\dh{S_{D7} = -N_7 T_7 \int\! d^8\xi \sqrt{-g_8} = -N_7 T_7 {\rm Vol}(S^4) \int\! d^3x ~d\psi  f^{1\over 2}r^4\sqrt{r^2 h +r'^2}  \sin^4 \psi~. }
The equations of motion work out to be
\eqn\di{ r'' + 4r' (1+{r'^2 \over B})\cot \psi  -{1 \over B}{dB \over dr} r'^2
-{1\over 2}{dB \over dr} -{(B+r'^2)\over A}{dA \over dr} =0 }
with
\eqn\dj{ A = r^4 f^{1\over 2}~,\quad B = r^2 h~. }
A solution in the large $r$ region is $x^9\equiv r\cos \psi =L$, which matches the
flat space D7-brane in \da.

Rather than taking the near horizon limit of the D3-brane solution, we have found it conceptually clearer to use the asymptotically flat version.   This makes it easier
to interpret the boundary conditions.  In particular, we will impose the asymptotic boundary condition $r\cos \psi = L$, and identify $L$ with the distance between
the branes at weak coupling, which in turn determines the mass of the fermions,
$m = L/2\pi \alpha'$.    In contrast, it is not entirely clear to us how to identify
this mass just using the near horizon probe solution.  This aspect of the problem
is different than for supersymmetric intersections, where $r\cos \psi =L$ can be
imposed as a boundary condition near the AdS boundary \refs{\KarchSH,\MMT,\KobayashiSB}.  In our case, the AdS
boundary behavior is more complicated.

\subsec{Zero temperature}

We first examine solutions in the zero temperature $(r_+=0)$ geometry.  It is
straightforward to solve \di\ numerically.    It is convenient to use the coordinate
$x^9 = r\cos \psi$, so that the boundary condition is
\eqn\dk{ x^9(\pi /2) = L~.}
The resulting solutions are displayed in Figure 3.
\fig{Zero temperature probe geometries (schematic). The repulsion of the probe D7-brane from the
D3-brane at the origin leads to spontaneous parity symmetry breaking and
dynamical mass generation.  The red line denotes the $x^9=0$ unstable symmetry preserving solution.    }{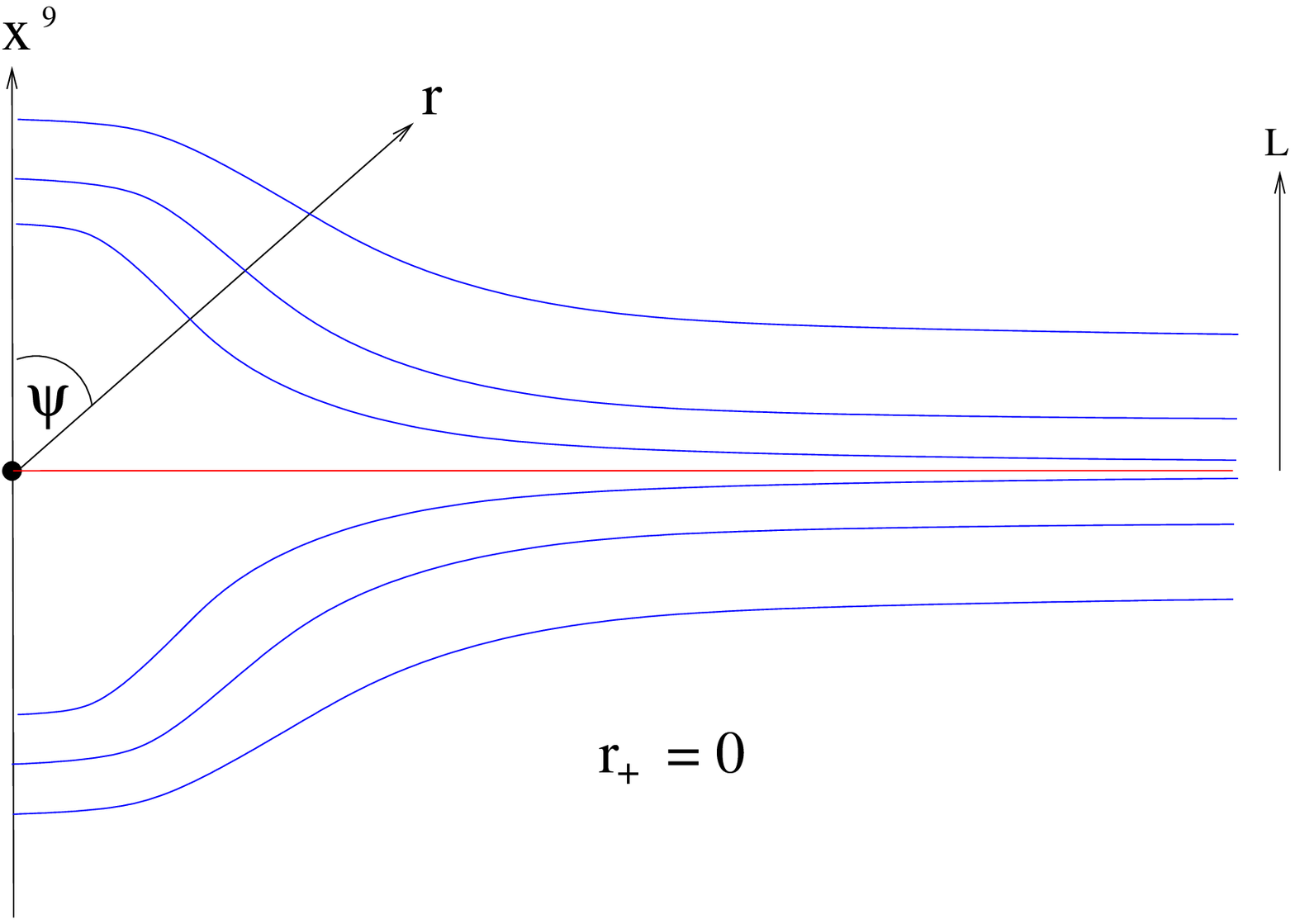}{3.5truein}

The main feature of note is that the probe is effectively repelled from the D3-branes, so that
$x^9(\psi=0)$ goes to a finite constant in the limit $L\rightarrow 0$.   This repulsion
arises from a competition of the positive and negative powers of $f$ appearing in \db,
and is familiar from the T-dual D0-D6 system \LifschytzIQ.

Precisely at $L=0$ there is a special solution with $x^9(\psi)=0$, corresponding to a
brane wrapping the equatorial $S^4$.   This solution is unstable to the brane slipping
off the equator.   To see this, note
that the probe metric in the small $r$ region is AdS$_4\times S^4$.   By expanding
out the Born-Infeld action, one finds that the slipping mode corresponds to a scalar
in AdS$_4$ with $m^2 R^2 = -4$, below the Breitenlohner-Freedman bound \BreitenlohnerBM\ of $m_{BF}^2 R^2 =-9/4$.   The endpoint of the instability is clear: the brane relaxes to one of
the $L=0$ solutions displayed in Figure 3.

The solutions nicely reproduce what we would expect for a theory of strongly coupled
fermions in $2+1$ dimensions, as was illustrated in the case of the Gross-Neveu
model with $g>g_c$.     For $L=0$ we see the spontaneous breaking of parity, which
takes $x^9\rightarrow -x^9$, and so interchanges the two stable solutions.    The $x^9=0$ solution is parity symmetric but unstable, as was the case in the field theory.
There is also a dynamically generated mass, as in the field theory.   The
string stretching between $x^9(\psi=0)$ and the D3-brane corresponds to a massive fermion. These solutions were also presented in \Rey.

For further comparison with the field theory we can compute the Chern-Simons
terms on the probe worldvolume.   In the presence of worldvolume gauge fields
the D7-brane action contains the Wess-Zumino term\foot{We write $\Tr A \wedge F$ as shorthand for the full non-Abelian Chern-Simons three form.}
\eqn\dl{ S_{WZ} = - {1\over 2}(2\pi \alpha' )^2 T_7 \int \! \Tr A \wedge F \wedge G_5  }
where $G_5$ is the RR 5-form field strength of the D3-brane background pulled back to
the worldvolume.    We first perform the integral over $S^4$ and $\psi$.   Now, the
D3-brane charge is quantized as
\eqn\dm{ \int_{S^5} G_5 = 2\kappa_{10}^2 T_3 N_3~.}
Our probe solutions have $\psi$ ranging from $0$ to $\pi/2$, corresponding to
one hemisphere of the $S^5$.  Thus the angular integration gives half the
result compared to \dm, with a plus or minus sign depending on whether we
wrap the north or south hemisphere.    Using
\eqn\dn { T_3 T_7 = {1 \over g^2 (2\pi)^4 (2\pi \alpha')^6}~,\quad
2\kappa_{10}^2 = g^2 (2\pi)^3(2\pi \alpha')^4~,}
we find the Chern-Simons term
\eqn\do{ S_{CS} = {k\over 4\pi}  \int\! d^3x ~ A\wedge F~,  }
with
\eqn\dpp{ k = \pm {1\over 2}N~,  \quad N=N_3 N_7~. }
We have now focused on the Abelian part of the full $U(N_7)$ gauge symmetry, and the factor of $N_7$ in $k$ arose from the trace. The effective action \do\ precisely matches the one-loop result in field theory, which is not surprising
since Chern-Simons terms do not arise beyond one-loop.

For the unstable $x^9=0$ solution we instead find $k=0$, since the pullback of
$G_5$ to the worldvolume vanishes in this case.

We now have all we need to describe the zero temperature quantum Hall plateau
transition.   In analogy to what we did on the field theory side, we consider
starting at positive $L$ and then smoothly evolving $L$ to negative values.
For $L\neq 0$ we can read off the conductivities from the worldvolume action.
As usual, the currents are defined by the variation of the on-shell action with
respect to the boundary values of the gauge fields.   In the present case we can
read off the answer without any detailed computation.  The longitudinal conductivity
is zero, as is always the case for a brane not passing through a horizon \refs{\KobayashiSB,\KarchPD}.  The
transverse conductivity arises from the  Chern-Simons term, giving $\sigma^T_{ij} = {k\over 2\pi}\epsilon_{ij}$.     Thus as we take $L$ towards zero we move along a
quantum Hall plateau, with fixed $\sigma^T$.  Precisely as we cross $L=0$ we see
that $k$ flips sign, and we jump from one plateau to another.   Here the jump is
$\Delta k = N_3 N_7$, but by instead moving just a single D7-brane we would have
$\Delta k = N_3$.

\subsec{Finite temperature}

\fig{Finite temperature probe embeddings (schematic):  At low temperatures ($r_+< r_+^{crit})$ the probe is described by a Minkowski embedding, and we jump from the upper branch of solutions to the lower
branch as  $L$ is lowered.      At high temperatures ($r_+> r_+^{crit}$) both Minkowski and
black hole embeddings exist.  As $L$ is lowered, we jump from a Minkowski embedding (denoted by the red dotted line) to a
black hole embedding.   Further reduction of $L$ takes us through a family of black hole embeddings, until
we eventually jump back to  the Minkowski embeddings. The transverse conductivity
varies continuously among the family of black hole embeddings.   }{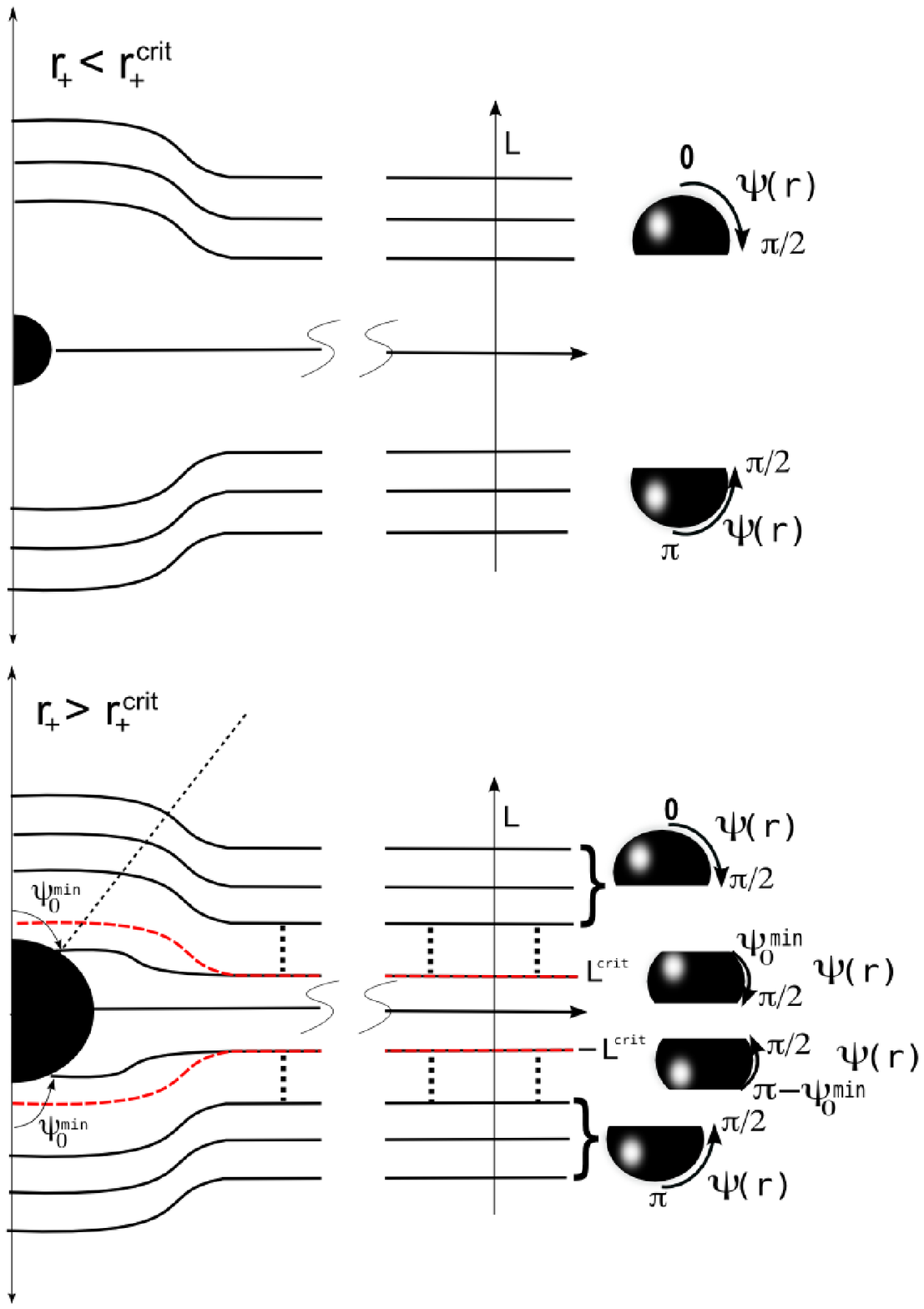}{4.0truein}

We now examine the finite temperature transition.    It is intuitively clear
that for $r_+ \ll R$ the situation is qualitatively the same as at $r_+=0$,
since there the  probe stayed a finite distance away from the origin.  New behavior
can only occur once  the black hole approaches the location of the probe.
One new
feature that can arise is that the probe can fall through the horizon.
Using standard parlance, we refer to probe geometries that stay outside the
horizon as ``Minkowski embeddings", and those that fall through as ``black hole
embeddings".    Black hole embeddings only exist\foot{For the special
case $x^9=0$ there is a black hole embedding. In fact, there are black hole
embeddings for other values of $L$ as well, but they are never energetically favored
so we ignore them.} for $r_+ > r_+^{crit}$, where $r_+^{crit}$ is
determined numerically to be $r_+^{crit} \approx .22$ in $R=1$ units.

In the low temperature regime, $r_+ < r_+^{crit}$, the situation matches what
we found on the field theory side.  As we lower $L$ we find  a
family of Minkowski embeddings.  At $L=0$ there are two solutions related by
parity, representing the two vacua with spontaneously broken parity.  As $L$ passes
through zero the probe undergoes a first order phase transition, jumping from
one vacuum to the other.  This represent the quantum Hall plateau transition, which
is thus discontinuous at low temperature.

Now turn to the  high temperature regime, $r_+ > r^{crit}_+$,  where there
are both Minkowski and black hole embeddings.  For sufficiently large (small) $L$ there
are only Minkowski (black hole) embeddings, while for an intermediate range both are
possible.   The Minkowski embeddings can be labeled by $x^9(\psi=0)$, and we
find that $L$ monotonically decreases with decreasing $x^9(\psi=0)$, reaching
a finite value for $x^9(\psi=0)=r_+$.
For the black hole embeddings we specify the angle $\psi_0$ at which they enter the horizon.   It turns out that $L(\psi_0)$ is a non-monotonic function of $\psi_0$, reaching a maximum at some finite $\psi_{0,crit}$; call this $L_{crit}$.   This leads to a discontinuous jump from a  Minkowski embedding to a black hole embedding:  starting from large
$L$ (and hence a Minkowski embedding) we lower $L$ until we reach $L_{crit}$, at
which point the probe jumps over to the black hole embedding with $\psi_0= \psi_{0,crit}$.\foot{This behavior is familiar from other examples of D-brane probes in black-brane backgrounds, see {\it e.g.} \MMT.}   From then on, further reduction of $L$ takes us smoothly through
a family of black hole embeddings.\foot{Due to the non-monotonic behavior noted above,  for fixed $L$ there are generically two such black hole embeddings, with distinct values of $\psi_0$.  However, the branch of solutions with the
larger $\psi_0$ is energetically preferred, and so we restrict attention to it.}      Eventually we pass smoothly through $L=0$
to negative $L$, at which point the story repeats itself in the reverse order.

The high temperature  quantum Hall plateau transition thus becomes partially smoothed out, while still retaining some discontinuous behavior.    To explore this further
we can compute the transverse conductivity as a function of $L$.   To do so
we return to \dl.   For the black hole embeddings $\psi$ takes the range
$\psi \in [\psi_0, {\pi\over 2}]$.    The probe thus wraps a fraction of the sphere
\eqn\dq{ F(\psi_0)=  {\int_{\psi_0}^{\pi \over 2} \!d\psi \sin^4 \psi \over \int_{-{\pi \over 2} }^{\pi \over 2} \!d\psi \sin^4 \psi} =-{1\over \pi} (\psi_0 - {\pi \over 2})+{1\over 3\pi} \sin \psi_0 \cos \psi_0 (5-2\cos^2 \psi_0)~. }
This is for $\psi_0 > {\pi \over 2}$; for $\psi_0 < {\pi \over 2}$ we instead have
$F(\psi_0) = F(\pi -\psi_0)$.
The coefficient of the Chern-Simons term is then
\eqn\dr{ k = \pm N F(\psi_0)~,}
where the $+ (-)$  sign corresponds to the upper (lower) hemisphere.   The result in
terms of $L$ can be obtained from the numerical solution for $\psi_0(L)$.

\fig{Transverse conductivities (schematic).  At low temperature ($r_+ < r_+^{crit}$) the conductivities jump
directly between the two plateaus.  At high temperature ($r_+ >r_+^{crit}$) the transition is
partially smoothed out.  The blue segments denote Minkowski embeddings; the green segment denotes
 black hole embeddings; and the red dot denotes the $x^9=0$ embedding.  The green segment starts
out with zero size at $r_+ = r_+^{crit}$, grows with increasing $r_+$, and reaches a limiting size
at large temperature.  The blue and green segments never connect, and so the plateau transition
is never completely continuous.    }{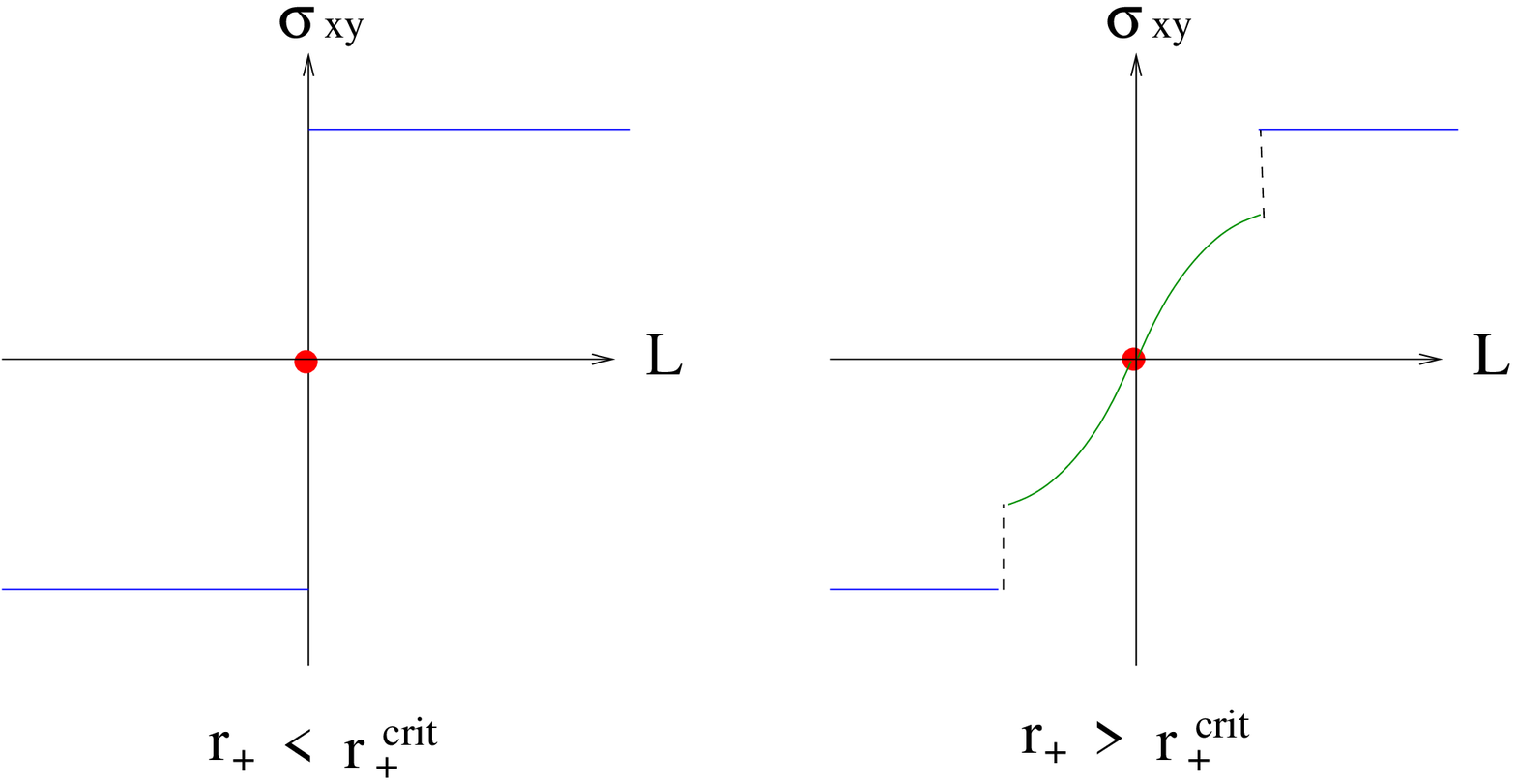}{4.5truein}

\newsec{Conductivity of black hole embedding}

Above, we found that at sufficiently high temperatures and for sufficiently light
fermions (i.e. small $L$)  the probe geometry  is described by a family of
black hole embeddings.   These solutions describe the transition region from
one quantum Hall plateau to the next.  It is of interest to compute the
DC conductivities during this transition.

To compute the conductivities on the gravity side we use the AdS/CFT correspondence.
In fact, since it is easy to do so, we will generalize the computation to include
a nonzero magnetic field and electric charge density.\foot{See \refs{\KarchPD,\OBannonIN,\AlbashBQ} for similar computations for the supersymmetric $3+1$ dimensional $D3-D7$ intersections and \BergmanSG\ for the Sakai-Sugimoto scenario.} We take the near horizon limit\foot{Strictly speaking, this involves taking $r_+ << R$ with errors of $O\left({r_+ \over R}\right)$. However as seen in the previous section, black hole embeddings only occur for $r_+> r_+^{crit} \approx .22 R$. Thus if we consider $r_+ \sim r_+^{crit}$ in this section, then we expect corrections of order $22\%$.} of the geometry, so that the metric on the probe
becomes the following
\eqn\eb{ ds_8^2 = {r^2 \over R^2}(-h dt^2+ dx^i dx^i)+{R^2 \over  r^2}(h^{-1}+r^2 \psi'^2)dr^2 +R^2 \sin^2 \psi d\Omega_4^2}
with
\eqn\ec{h =1-{r_+^4 \over r^4}~. }
Note that we are parameterizing the probe via $\psi = \psi(r)$.
Following the logic of \BhattacharyyaJC\ it is most convenient to work in
Eddington-Finkelstein coordinates, since they extend smoothly across the
future horizon.  Defining $\rt(r)$ via
\eqn\eca{ \rt = \int^r_{r_+} \! dx \sqrt{1+h(x)x^2\psi'(x)^2} }
and defining the advanced coordinate $v$ in the usual fashion,  the metric takes the form
\eqn\ybb{ ds^2= 2dvd\rt -U(r)dv^2 +{r^2\over R^2}dx^i dx^i +R^2 \sin^2 \psi d\Omega_4^2}
with
\eqn\ybc{ U = {r^2\over R^2} h = {r^2\over R^2} -{r_+^4 \over R^2 r^2}~.}

The action of the probe is given by the Born-Infeld action plus  the Chern-Simons
term with coefficient \dr.    After integrating over the $S^4$ the Born-Infeld contribution to the
probe action becomes
\eqn\ed{ S_{BI}= -\tau \int\! d^4 x \sin^4 \psi \sqrt{-\det(g_4 +2\pi \ap F)}}
with
\eqn\ee{ \tau =N_7  T_7 R^4 V(S^4) =N_7 \left({1\over g_s (2\pi)^7 \alpha'^4}\right)(4\pi g_s \alpha'^2 N_3)({8\over 3 }\pi^2)={1 \over 3\pi^2} {1\over (2\pi \alpha')^2}N_3 N_7~.}

The Chern-Simons term can be written
\eqn\eea{S_{CS}= {1\over 4\pi} \int\! ~ d f(\rt)\wedge A\wedge F }
The function $f(\rt)$ is determined by the pullback of the background flux onto the
brane worldvolume, and depends on the choice of embedding.     The Chern-Simon coefficient in the boundary theory is obtained
by integrating from the horizon out to the boundary, $k = \int_{\rt_+}^\infty \! d\rt  \p_\rt f(\rt)  $.

We look for a solution of the gauge field equations of motion, subject to the
boundary condition of  nonzero electric and magnetic fields on the boundary.
We thus consider the  gauge field ansatz
\eqn\ef{\eqalign{A_{\rt} &= 0 \cr  A_v& = a_v(\rt)+E^i x^i \cr A_j &= {1 \over 2} B \epsilon_{ij}x^i +a_j(\rt)~. }}
We raise and lower $ij$ indices with $\delta_{ij}$.

Working out the field strengths and plugging into \ed\ gives
the action\foot{To avoid clutter we now absorb factors of  $(2\pi \alpha')$ into the
field strengths, and will restore them later.}
\eqn\eg{\eqalign{ S_{BI}= -\tau \int\!d^4x &\sin^4 \psi \Big[ (B^2+{r^4\over R^4})(1- (\p_{\rt} a_v)^2) +U{r^2\over R^2}(\p_{\rt} a_i)^2 \cr & -2{r^2\over R^2} E^i \p_{\rt} a_i +2B \p_{\rt} a_v \epsilon_{ij} E^i \p_{\rt} a_j -(\epsilon_{ij} E^i \p_{\rt} a_j)^2 \Big]^{1\over 2}~,   }}
and (omitting the $a_\mu$ independent part)
\eqn\ega{S_{CS} = {1\over 2\pi} \int\! d^4x \p_{\rt}f(\rt) \Big[ B a_v(\rt)+\epsilon_{ij}
a_i(\rt) E_j\Big]~.}
In working out $S_{CS}$ we first allowed $a_\mu$ to depend on $x^i$, integrated by parts
those terms involving $\p_i a_\mu$, and then removed the $x^i$ dependence; this is
necessary in order to get the correct equations of motion. Alternatively, one can just insert by hand an extra factor of two to make up for the fact that the full CS term is quadratic in gauge fields while the Lagrangian evaluated on our ansatz is linear in the fluctuations $a_{v , j}$.

We now define the following $\rt$ dependent charge and currents
\eqn\eh{\rho(\rt) = {\p {\cal L} \over \p (\p_\rt a_v)}~,\quad j^i(\rt) =   {\p {\cal L} \over \p (\p_\rt a_i)}~.}
From the AdS/CFT dictionary the charge density and current of the boundary theory are given by the boundary values of these functions \OBannonIN\
\eqn\ehb{ \rho \equiv \rho(\rt)|_{\rt=\infty}~,\quad j_i \equiv j_i(\rt)|_{\rt=\infty}~.}
Notice that the Chern-Simons term doesn't contribute directly to these expressions, since in \ega\ there are no radial derivatives of the gauge field components.

Now, $\rho(\rt)$ and $j_i(\rt)$ depend
on the radial coordinate $\rt$ according to the $a_{v,i}$ equations of motion:
\eqn\eha{ \p_\rt \rho(\rt) = {B\over 2\pi} \p_\rt f(\rt)~,\quad \p_\rt j_i(\rt) ={1\over 2\pi}\p_\rt f(\rt) \epsilon_{ij}E_j.}
Thus, relative to the horizon ($\rt=\rt_+)$  these functions measured at infinity are shifted by an amount proportional to the Chern-Simons coefficient,
\eqn\ehb{ \rho = \rho(\rt_+)+{kB\over 2\pi}~,\quad  j_i = j_i(\rt_+)+{k\over 2\pi}\epsilon_{ij}E_j~.}

\def\rhoh{\hat{\rho}}

We will work perturbatively in the electric field $E^i$. For $E^i=0$, by inverting \eh\ we have
the solution
\eqn\ei{ \p_{\rt} a_v = { \rho(\rt)/\tau \over \sqrt{\rho(\rt)^2/\tau^2 +(B^2 +r^4/R^4)\sin^8 \psi}}}
and with $a^i=0$.

Now we go to first order in $E^i$.   To this order we can take $a_v$ as in \ei.
We then write out the expression for the current in \eh.  It is most convenient
to write the result at $\rt_+$, where $U=0$ and $\psi = \psi_0$, which gives
\eqn\ej{ j^i(\rt_+) = \tau{\sqrt{\rho(\rt_+)^2/\tau^2 +(B^2 +r_+^4/R^4)\sin^8 \psi_0} \over B^2 +r_+^4/R^4}({r_+^2\over R^2} E^i + B \p_{\rt} a_v \epsilon_{ij} E^j )~.  }
Writing
\eqn\ek{ j^i = \sigma^L E^i  + \sigma^T \epsilon_{ij}E^j }
we get, after restoring the factors $2\pi \alpha'$ and using \ee\ and \ehb,
\eqn\el{\eqalign{ \sigma^L & =  {N \over 3\pi^2} {r_+^2\over R^2 }{\sqrt{(2\pi \alpha' \rhoh)^2 /(2\pi\alpha')^4\tau^2  +\left((2\pi \alpha'B)^2 +r_+^4/R^4\right)\sin^8 \psi_0} \over (2\pi \alpha'B)^2 +r_+^4/R^4} \cr \sigma^T & =    {(2\pi \alpha'\rhoh)(2\pi \alpha'B) \over (2\pi \alpha'B)^2 +r_+^4/R^4}+{k\over 2\pi}~, }}
where we defined the shifted charge (equivalent to $\rho(\rt_+)$)
\eqn\ela{\rhoh = \rho -{kB\over 2\pi}~.}
Note that we do not have to add in the Chern-Simons term to this result; it is already
taken into account.

Our basic plateau transition occurs for $\rho=B=0$.  At fixed temperature, as we evolve
through the one parameter family of Minkowski and black hole embeddings we obtain the
conductivities  illustrated in Fig. 6.

\fig{Longitudinal and transverse conductivities (schematic).    }{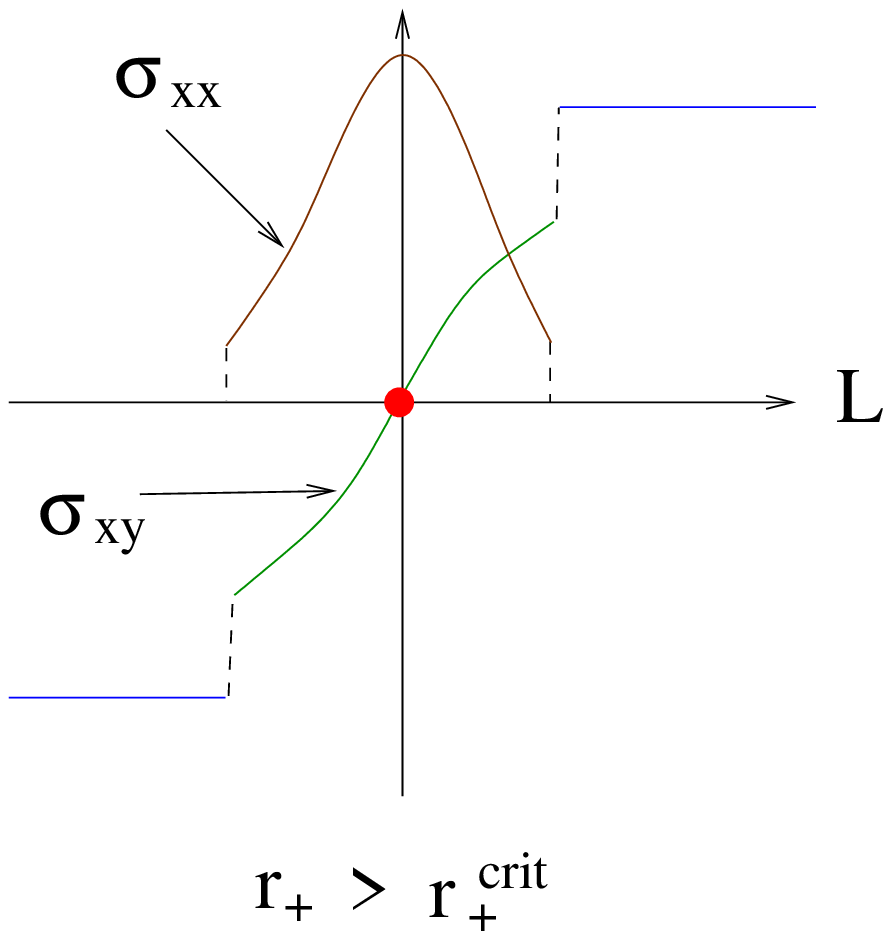}{2.5truein}

To make a heuristic comparison with the free fermion result we set $\rho=B=0$, which gives
\eqn\em{ \sigma^T=0~,\quad\sigma^L={N \over  3\pi^2}~, \quad\quad (\rho=B=0) ~.}
We see that the longitudinal conductivity at strong coupling differs from the free fermion result by  a factor $3\pi^2 / 16 \approx 1.85$.   We call this comparison heuristic because we are actually comparing
two different things.  In general the AC conductivity is a function of $\omega/T$; in the free fermion
case we first set $T=0$ and then $\omega=0$, while on the gravity side we first set $\omega=0$ and then $T=0$.  Thus
the comparison made here is really between $\omega/T=\infty$ and $\omega/T=0$, which can give different
results.

Another relevant limit is zero temperature but finite charge and magnetic field:
\eqn\en{ \sigma^T ={\rho\over B}~,\quad  \sigma^L=0~,\quad\quad  (r_+=0).}
This gives the standard expression for the classical Hall conductivity.  As is well
known, this result follows from boost invariance.  Note that in  \el\ the
Hall conductivity is a nontrivial function of temperature.  This occurs because
the finite temperature black hole background is not boost invariant.  From the
dual field theory point of view, we are computing the conductivity with respect
to the 3-7 strings in the background of a thermal gas of 3-3 strings.  The electric
field accelerates the 3-7 strings but not the 3-3 strings. See \refs{\KarchPD,\OBannonIN} for more discussion of this point.

Finally, let us discuss the possibility of defining the critical exponent $1/\nu z$.  Strictly speaking,
this is not well defined here, since the definition of critical exponents is based on having scaling behavior
at the transition, which is not realized here due to the discontinuity in the conductivity.   Nevertheless,
one could contemplate defining an exponent in terms of the slope of the curve.  The slope at the origin
is ill-defined at the transition point, since the continuous part of the curve has zero size at the
critical temperature.   Alternatively, one could compute an average slope $\left({\Delta \sigma_{xy} \over \Delta L}\right)$ where $\Delta \sigma_{xy}$ is the difference in transverse conductivity between the two plateaus and $\Delta L$ is some estimated width of the transition region. Due to the instability of the numerics in the critical temperature region, this is a rather difficult computation which we hope to return to in future work.

\newsec{Discussion}

We have shown that quantum Hall plateau transitions can be realized in string theory
in terms of the relative motion of D3 and D7 branes.  The strongly coupled version
can be studied using techniques of gauge/gravity duality.  We found that the details
of the transition share some similarities and some differences with the usual transition
realized experimentally.  Strong coupling triggers dynamical symmetry breaking and
mass generation, which leads to a transition that remains discontinuous at low temperature.
At high temperature it is partially smoothed out, but  a finite jump in conductivities persists.
The relation between the two picture is summarized by comparing Figures 1 and 6.

There are a number of natural generalizations of the current setup that we hope to report on in the near future, including adding nonzero magnetic fields and charge densities. There are
other extensions that would be interesting to explore, if possible.  One would be to
realize a fractional QHE plateau transition.  Our present theory has a weak coupling description in terms of D-branes in Minkowski space, and hence has a purely integer charge
spectrum.  To obtain the fractional case, any weak coupling limit would have to exhibit
charge fractionalization, possibly due to an orbifold geometry.   It would of course
be interesting to find a setup in which the plateau transitions corresponds to a quantum
phase transition, rather than the partially discontinuous finite temperature transition
obtained here.   Much of the current interest in the QHE effect involves the study of
graphene, and one can hope to make contact with that work, as in \Rey.

\bigskip
\noindent {\bf Acknowledgments:} \medskip \noindent We thank Ivailo Dimov, Clifford Johnson,
Pallab Goswami, Esko Keski-Vakkuri, and Gordon Semenoff  for discussions, and Soo-Jong Rey for communication.     Work of PK and JD  supported in part by NSF grant PHY-0456200.

\listrefs
\end